\documentclass[prd,aps,eqsecnum,showpacs,nofootinbib]{revtex4}
\usepackage{epsfig,graphicx,latexsym}
\def\rmd{{\rm d}}
\def\msun{$M_{\odot}$}
\def\MNRAS{{\em Mon. Not. Roy. Astron. Soc. }}
\def\CQG{{\em Class. Quantum Grav. }}
\def\PRD{{\em Phys. Rev. D }}
\def\apj{{\em Astrophys. J.}}
\newcommand{\erf}[1]{(\ref{#1})}

\begin{document}

\title[Detecting LISA EMRI events using HACR]{Detecting extreme mass ratio inspiral events in LISA data using the Hierarchical Algorithm for Clusters and Ridges (HACR)}
\author{Jonathan Gair\footnote{email address: jgair@ast.cam.ac.uk}}
\affiliation{Institute of Astronomy, University of Cambridge, Madingley 
Road, Cambridge, CB3 0HA, UK}
\author{Gareth Jones}
\affiliation{Cardiff School of Physics and Astronomy, Cardiff University, Queens Buildings,
The Parade, Cardiff, CF24 3AA, UK}

\date{\today}

\begin{abstract}
One of the most exciting prospects for the Laser Interferometer Space Antenna (LISA) is the detection of gravitational waves from the inspirals of stellar-mass compact objects into supermassive black holes. Detection of these sources is an extremely challenging computational problem due to the large parameter space and low amplitude of the signals. However, recent work has suggested that the nearest extreme mass ratio inspiral (EMRI) events will be sufficiently loud that they might be detected using computationally cheap, template-free techniques, such as a time-frequency analysis. In this paper, we examine a particular time-frequency algorithm, the Hierarchical Algorithm for Clusters and Ridges (HACR). This algorithm searches for clusters in a power map and uses the properties of those clusters to identify signals in the data. We find that HACR applied to the raw spectrogram performs poorly, but when the data is binned during the construction of the spectrogram, the algorithm can detect typical EMRI events at distances of up to $\sim2.6$Gpc. This is a little further than the simple Excess Power method that has been considered previously. We discuss the HACR algorithm, including tuning for single and multiple sources, and illustrate its performance for detection of typical EMRI events, and other likely LISA sources, such as white dwarf binaries and supermassive black hole mergers. We also discuss how HACR cluster properties could be used for parameter extraction.
\end{abstract}

\maketitle

\section{Introduction}
\label{intro}
Astronomical observations indicate that many galaxies host a supermassive black hole (SMBH) in their centre, which is surrounded by a cluster of stars. Stellar encounters in the cluster modify the stellar orbits over time and can put objects onto trajectories that pass very close to the SMBH. If this happens, the star loses energy and angular momentum to a burst of gravitational radiation emitted near periapse, and this may leave the object bound to the central black hole. Gravitational radiation dominates the subsequent evolution, and the object gradually inspirals into the SMBH. If the captured star is a compact object, i.e., a white dwarf, neutron star or black hole (so that it is not tidally disrupted) and the SMBH has mass
$M\sim\mbox{few}\times10^{5}M_{\odot}$--$10^{7}M_{\odot}$, the gravitational waves (GWs) emitted during the last several years of inspiral will be at frequencies that the planned space-based gravitational wave (GW) observatory LISA will be able to detect. 

The rate at which these extreme mass ratio inspiral (EMRI) events occur in the Universe is highly uncertain, but is likely to be at most only a few times per year in each Gpc$^3$ of space \cite{freitag01,jon04}. LISA EMRI events are thus unlikely to be closer than $~1$Gpc, at which distance the typical instantaneous strain amplitude is $h\sim 10^{-22}$. This must be compared to the characteristic noise strain in the LISA detector of $\sim 5 \times 10^{-21}$ at the floor of the noise curve near 5 mHz \cite{curt98,leor04}. EMRI waveforms will be detectable for several years before plunge, which makes detection possible by building up the signal-to-noise ratio over many waveform cycles using matched filtering. However, matched filtering involves searching a bank of templates that describe all possible signals that might be present in the data. An EMRI waveform depends on $17$ parameters (although several of these are not important for determining the waveform phasing) and LISA will detect up to $\sim10^5$ cycles of the waveform prior to plunge. Estimating that one template might be required per cycle in each parameter, and $\sim 6$ important parameters, gives an estimate of $10^{30}$ templates required for the simplest case of a search for a single EMRI embedded in pure Gaussian noise. This is far more than can be searched in a reasonable computational time \cite{jon04}.

Analysis of LISA data is further complicated by the fact that it is signal dominated, i.e., at any moment the data stream includes not only instrumental noise but thousands of signals of different types which overlap in time and frequency. The optimal matched filter should therefore be a superposition of all the signals that are present. Techniques exist to construct such a global matched filter iteratively, such as Markov Chain Monte Carlo (MCMC) methods, and are currently being investigated in the context of LISA \cite{cornish05,umstatter05,wickham06,cornishporter06}, including for characterisation of LISA EMRIs \cite{stroeer06emri}. However, the MCMC approach still relies on matched filtering. Although this is done in an efficient way that typically requires only $\sim 10^7$ waveform evaluations \cite{stroeer06emri}, these $10^7$ templates need to be either generated on the fly, or looked up in a template bank. For EMRIs, the computational cost of either approach may be prohibitively high, unless some advance estimate has been made of the parameters of the signals present in the data. To devise such parameter estimation techniques, it is reasonable to first consider the problem of detecting a single source in noisy data, before using and adapting the methods to the case of multiple sources. It is this second problem, searching for a single source while ignoring source confusion, that work on EMRI searches has concentrated on so far.

One possible algorithm for EMRI detection is a semi-coherent approach --- rather than search for the full waveform, begin with a coherent search for $\sim2$--$3$ week segments of EMRI waveforms, before following up with a second stage where the power in each segment is summed up through sequences of segments that correspond to inspirals. Assuming reasonable computational resources, this technique could detect individual EMRI events out to a redshift $z \approx 1$ \cite{jon04}, which would mean as many as several hundred EMRI detections over the duration of the LISA mission, although this is very dependent on the intrinsic astrophysical rate of EMRI events. The semi-coherent method, although computationally feasible, makes heavy use of computing resources. However, the high potential event rate suggests that it might be possible to detect the loudest several EMRI events using much simpler, template-free techniques, at a tiny fraction of the computational cost.

A promising technique for the detection of EMRIs, and other types of LISA sources, is a time-frequency analysis --- divide the LISA data stream into shorter segments and construct a Fourier spectrum of each, hence creating a time-frequency spectrogram of the data, and then search this time-frequency map for features. The simplest possible time-frequency algorithm is an Excess Power search, i.e., a search for unusually bright pixels in the spectrogram.  While this performs poorly when applied to the raw data, if the data is binned first, the Excess Power method is able to detect typical EMRI events at distances of up to $\sim 2.25$Gpc \cite{wengair05,gairwen05}, or about half the distance of the semi-coherent search \cite{jon04}. The disadvantage of the Excess Power method in isolation is that it does not provide much information about the source parameters, but merely indicates that a source is present in the data. A follow up analysis must therefore be used to extract information about events identified by the Excess Power search \cite{wen06}.

In this paper we consider a somewhat more sophisticated time-frequency algorithm, the Hierarchical Algorithm for Clusters and Ridges (HACR) \cite{heng04}. This method involves first identifying unusually bright pixels in the time-frequency map, then constructing a cluster of bright pixels around it, before finally using the number of pixels in the cluster as a threshold to distinguish signals from noise events. The properties of the HACR clusters encode information about the source, and thus in a single analysis HACR allows both detection and parameter estimation. 

We have found that when HACR is applied to the unbinned spectrogram, it performs poorly, but if the spectrogram is first binned via the same technique used for the Excess Power search \cite{wengair05,gairwen05}, we find that HACR outperforms the Excess Power search, as we would expect. HACR is able to detect typical EMRI events at distances of $\sim 2.6$Gpc, which is a little further than the Excess Power technique. However, the HACR clusters associated with detection events tend to have several hundred pixels, and thus encode a significant amount of information about the source. The HACR search can be tuned to be sensitive to a specific source at a specific distance, or to a specific source at an unknown distance, or to an unknown source at an unknown distance. While the detection performance for a specific source does depend on how the HACR thresholds are tuned, we find that the variation of detection rate is not huge and so a single HACR search could be used to detect multiple types of event in a search of the LISA data. The HACR search encompasses the Excess Power search as a subset (with the pixel threshold set to 1), which will allow us to compare HACR's performance to the performance of the Excess Power algorithm in this paper.

The paper is organised as follows. In Section~\ref{LISA} we describe LISA, the LISA data stream and LISA sources, including the waveform models we use for source injections and the noise model we use to estimate LISA's noise spectral density. In Section~\ref{HACR} we introduce the HACR algorithm, discuss how the time-frequency map may be binned to improve detection of signals and describe how the HACR search can be characterised when no signals are present. In Section~\ref{EMRIperformance}, we describe how the HACR thresholds may be tuned for detection of a single source or multiple sources, while maintaining a constant overall false alarm probability, and we evaluate HACR's performance at detecting a range of EMRI signals in LISA data. Although EMRIs are the main focus of our analysis, time-frequency techniques like HACR may play a role in the detection and identification of other sources in the LISA data stream. For this reason, in Section~\ref{otherperformance} we evaluate HACR's performance at detecting other classes of signal, specifically the gravitational waves from white dwarf binaries and the mergers of supermassive black hole binaries. In Section~\ref{paramest} we will briefly describe how the properties of the event candidates identified by HACR could be used to estimate parameters of the source, although a more detailed investigation of this is reserved for a future paper. Finally, in Section~\ref{conclusion} we summarise our results.

\section{LISA overview --- detector and sources}
\label{LISA}
\subsection{The LISA mission}
\label{LISAmission}
The LISA detector will consist of three spacecraft in heliocentric Earth-trailing orbits, $5$ million kilometres apart at the corners of an (approximately) equilateral triangle (see \cite{LISAppa} for a full description of the mission). There will be two lasers running between each pair of spacecraft, one in each direction, and it is the differences in laser phase between the various light travel paths that indicate that gravitational waves are passing through the detector. In the raw data, the laser phase difference is totally dominated by laser frequency noise. However, this can be suppressed without eradicating the GW signal using Time Delay Interferometry (TDI, see for instance \cite{vallis05} and references therein). At high frequencies, there are three independent TDI channels in which the noise is uncorrelated, which are typically denoted $A$, $E$ and $T$. At low frequencies there are essentially only two independent data channels, since LISA can be regarded as a superposition of two static orthogonal Michelson interferometers over relevant timescales. These two low frequency response functions, denoted $h_{I}$ and $h_{II}$, are defined in \cite{curt98}. In this analysis we treat the LISA data stream as consisting of only the latter two channels, since our sources are at comparatively low frequency, and the Michelson responses are quick and easy to compute. While not a totally accurate representation of LISA, this approach incorporates the modulations due to the detector motion in a reasonable way and so is sufficient for the qualitative nature of the current analysis.

LISA is expected to detect gravitational waves from many sources of several different types. LISA should detect many millions of compact object binaries (white dwarf -- white dwarf, white dwarf -- neutron star, neutron star -- neutron star) in our own galaxy, which generate essentially monochromatic gravitational wave signals (up to detector modulations). These binary signals are sufficiently dense in frequency space that the majority will not be individually resolvable, but form a confusion foreground from which other sources must be extracted. However, several thousand binaries will still be resolvable \cite{LISAppa}. LISA should also detect several (estimates suggest as many as a few tens per year, e.g., \cite{sesana05}) signals from the final inspiral and merger of SMBH binaries. For SMBHs of appropriate mass, $M\sim 10^{4}M_{\odot}$--$10^{7}M_{\odot}$, these mergers will be visible out to very high redshifts and will appear in the LISA data stream with very high signal-to-noise ratio. Over the course of its planned three year mission, LISA should also detect as many as several hundred EMRI events as discussed in the introduction, and may detect GW bursts and perhaps a stochastic GW background.

\subsection{A typical EMRI source}
\label{sourcemodel}
In this paper we concentrate on the issue of detection of EMRI events and to do so we must consider a typical EMRI signal. Work on the semi-coherent search suggested that the LISA EMRI event rate would be dominated by the inspiral of black holes of mass $m \sim 10M_{\odot}$ into SMBHs with mass $M\sim 10^6 M_{\odot}$ \cite{jon04}. An EMRI will be detectable for the last several years of the inspiral, and hence could last for a significant fraction of the LISA mission duration. Moreover, since the EMRI will typically be captured with very high eccentricity and random inclination with respect to the equatorial plane of the SMBH, the inspiral orbit is likely to have some residual eccentricity and inclination at plunge. Theoretical models \cite{volon05} and some observational evidence \cite{miniutti04,fabian05} indicate that most astrophysical black holes will have significant spins. Bearing all this in mind, we choose as a ``typical'' EMRI event (which we shall refer to as source ``A'') the inspiral of a $10 M_{\odot}$ black hole into a $10^6 M_{\odot}$ SMBH with spin $a=0.8 M$. We assume conservatively that the LISA mission will last only three years ($3 \times 2^{25}$s) and that the EMRI event is observed for the whole of the LISA mission, but plunges shortly after the end of the observation. This sets the initial orbital pericentre to be at $r_p \approx 11M$. We take the eccentricity and orbital inclination at the start of the observation to be $e = 0.4$ and $\iota = 45^{{\rm o}}$ and fix the sky position in ecliptic coordinates to be $\cos\theta_S = 0.5$, $\phi_S = 1$. The orientation of the SMBH spin is chosen such that if the SMBH was at the Solar System Barycentre, the spin would point towards ecliptic coordinates $\cos(\theta_K)=-0.5$, $\phi_K=4$. These latter orientation angles were chosen arbitrarily, but are non-special. We generate the EMRI waveform using the approximate, ``kludge'', approach described in \cite{kludgepaper,GG06}. These kludge waveforms are much quicker to generate than accurate perturbative waveforms, but capture all the main features of true EMRI waveforms and show remarkable faithfulness when compared to more accurate waveforms. In addition to source ``A'', we will consider two other EMRI injections. These have the same parameters as ``A'', except for the initial orbital eccentricity, which is taken to be $e=0$ for source ``K'' and $e=0.7$ for source ``N''. The waveforms and waveform labels used are the same as those examined in the context of the Excess Power search in \cite{gairwen05}, to facilitate comparison.

In Section~\ref{otherperformance}, we will examine the performance of HACR in detecting other LISA sources, namely white dwarf binaries and SMBH mergers. For both of these sources, we take the waveform model, including detector modulations, from \cite{curt98}. Although more sophisticated SMBH merger models are available, the prescription in \cite{curt98} is sufficiently accurate for our purposes. The waveform model for a non-evolving white dwarf binary is very simple and has been well understood for many years and is summarised in \cite{curt98}.

\subsection{Modelling LISA's noise spectral density}
To characterize the search, we need to include the effects of detector noise. To do this, we use the noise model from Barack and Cutler \cite{leor04}\footnote{NB The published version of this paper contains an error in the expression for $S_h$, which has been corrected in the preprint gr-qc/0310125. We use the corrected expressions here.}, which includes both instrumental noise and ``confusion noise'' from the unresolvable white dwarf binary foreground. The noise spectral density is given by
\begin{eqnarray}
&& S_h(f) = {\rm min} \left\{S_h^{inst}(f) \, \exp\left(\kappa T^{-1}_{mission}\,\rmd N/\rmd f\right), S_h^{inst}(f) + S_h^{gal}(f) \right\} + S_h^{ex. gal}(f) \\
\mbox{where } \, && S_h^{inst}(f) = 9.18\times10^{-52} \,\, f^{-4} + 1.59 \times 10^{-41} + 9.18 \times 10^{-38} \,\,f^2 {\rm Hz}^{-1} \\
\mbox{and }&& S_h^{gal}(f) = 50 \, S_h^{ex. gal}(f) = 2.1 \times 10^{-45} \left(\frac{f}{1\,{\rm Hz}} \right)^{-7/3} \,\, {\rm Hz}^{-1}.
\label{LISAsh}
\end{eqnarray}
In this, the parameter $\kappa\,T^{-1}_{mission}$ measures how well white dwarfs of similar frequency can be distinguished, and we take $\kappa\,T^{-1}_{mission} = 1.5/{\rm yr}$ as in \cite{leor04}. In practice, rather than adding coloured noise to the injected signal, we first whiten the signal using this theoretical noise prescription and then inject it into white Gaussian noise. The procedures are equivalent, under the assumption that the LISA data stream can be regarded as stationary and supposing that the noise spectral density is known or can be determined. This is likely to be a poor assumption, but a more accurate analysis is difficult and beyond the scope of this paper.

\section{The Hierarchical Algorithm for Clusters and Ridges}
\label{HACR}

\subsection{Description of method}
\begin{figure}[!hb]
\begin{center}
\includegraphics[height=3.5in]{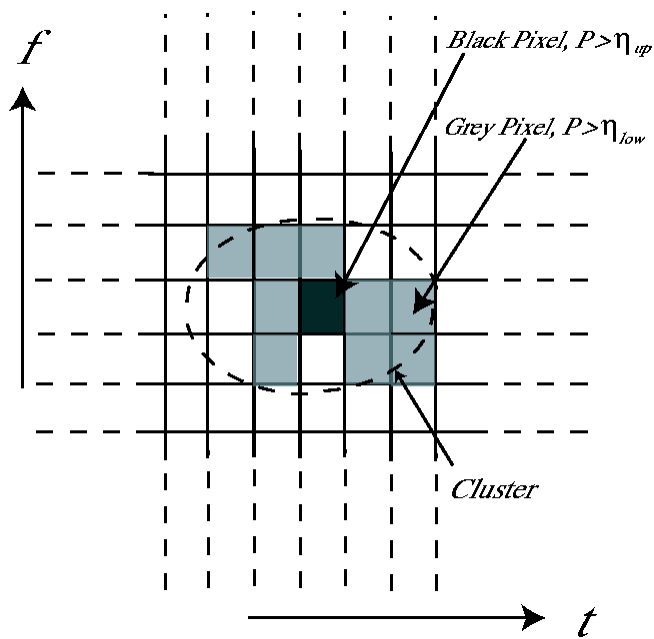}
\end{center}
\caption{A time-frequency map illustrating the HACR algorithm. Pixels
with power $P_{i,j} > \eta_{up}$ are classified as {\em black pixels}. 
Surrounding pixels with $P_{i,j} > \eta_{low}$ are then classified as
{\em grey pixels} building a cluster around the black pixel. The 
cluster is classified as an event candidate if the number of pixels it
contains, $N_{c}$, exceeds the threshold $N_{p}.$}
\label{HACRIll}
\end{figure}

The HACR algorithm identifies clusters of pixels containing excess power in a time-frequency map and represents a variation of the TFClusters algorithm \cite{jsylvestre}. In a given time-frequency map, we denote the power in a pixel as $P_{i,j}$ where $i$ and $j$ are the time and frequency co-ordinates of the pixel. HACR employs two power thresholds, $\eta_{up} > \eta_{low}$ and a pixel threshold, $N_p$. At the first stage, the algorithm identifies all {\it black pixels} with $P_{i,j} > \eta_{up}$ and all {\it grey pixels} with $P_{i,j} > \eta_{low}$. At the second stage, HACR takes each black pixel in turn and counts all the grey pixels that are connected to that black pixel through a path
of touching grey pixels. Touching is defined as sharing an edge or corner. This process is repeated for each black pixel. To be classified as an {\it event candidate} a cluster of pixels must have
$N_{c} > N_{p}$ where $N_{c}$ is the number of pixels contained
in a particular cluster. The algorithm is illustrated in Figure~\ref{HACRIll}

There is some degeneracy between the thresholds, particularly $\eta_{low}$ and $N_p$. Choosing a low value of $\eta_{low}$ tends to make clusters larger, but this can be partially compensated for by using a larger value for the pixel threshold, $N_p$. In our first analysis, we fixed $\eta_{low}$ and tuned only $N_p$ for this reason. However, tuning $\eta_{low}$ as well can enhance the detection rate by $10-15\%$. Results in this paper use tuning over both thresholds. The thresholds affect not only the detection rate, but also parameter extraction. A smaller $\eta_{low}$ tends to make clusters larger. This might increase the detection rate, but it also increases the number of noise pixels in each detected cluster, hampering parameter extraction. The optimal thresholds for the final search will ultimately come from a compromise between greater reach and more accurate parameter extraction. In a future paper, when we explore parameter estimation, we will examine this issue more carefully. In the current paper, we look only at maximizing the detection rate.

\subsubsection{Investigating ``binning'' of the time-frequency maps}
It is possible to improve the performance of the search by ``binning''
the time-frequency maps. This binning procedure was the key stage in 
the simple Excess Power search discussed in \cite{wengair05,gairwen05}.

This binning procedure involves constructing an average power map using boxes of particular size. The average power contained within a box is defined by
\begin{equation}
P_{i,j}^{n,l}= \frac{1}{m}\,\sum^{n-1}_{a=0} \sum^{l-1}_{b=0} P_{i+a,j+b}
\end{equation}
where $n$, $l$ are the lengths of the box edges in the time and frequency dimensions respectively and $m=n\times l$ is the number of data points
enclosed. This average power is computed for a box aligned on {\it each} pixel in the time-frequency map. Adjacent pixels in the average power map are therefore not independent. In practice, for ease of computation we choose the alignment so that the pixel is in the top left hand corner of the box.
As in \cite{wengair05}, we use only box sizes $(n,l)=(2^{n_t},2^{n_f})$ for all possible integer values of $n_t$ and $n_f$. We denote the total number of different box shapes used as $N_{box}$.

For a given source, the box size that will do best for detection will be large enough to include much of the signal power but small enough to avoid too much noise contribution. This optimum will be source specific due to the wide variation in EMRI waveforms. The inspiral of a $0.6$\msun white dwarf occurs much more slowly than that of a $10$\msun black hole, so in the first case, the optimal box size is likely to be longer in the time dimension. GWs from an inspiral into a rapidly spinning black hole or from a highly eccentric inspiral orbit are characterized by many frequency harmonics, often close together. In that case, a box that is wider in the frequency dimension may perform well. In designing a search, a balance must therefore be struck between having sensitivity to a range of sources and increasing the reach of the search for a specific source. We will consider this more carefully in Section~\ref{targsearch}.

\subsubsection{Efficient ``binning'' method} 
\label{effbinning}
\begin{figure}[!hb]
\begin{center}
\includegraphics[height=3.5in,angle=0]{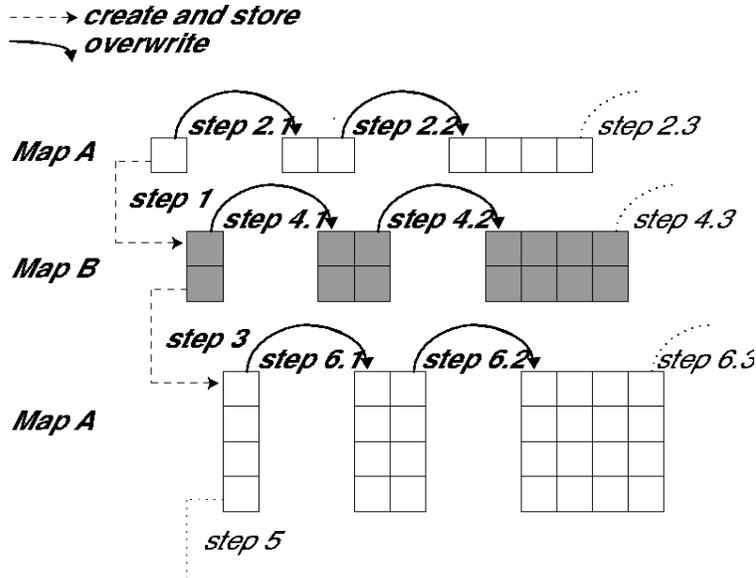}
\end{center}
\caption{Schematic showing the efficient binning of the time-frequency map, following the algorithm described in the text.}
\label{binning}
\end{figure}

The binned spectrograms for each box size can be generated in a particular sequence that 
improves the efficiency and speed of the search as shown in 
Figure~\ref{binning}.  
We first construct the unbinned $(n=1,l=1)$ map of the data and store it as map A.
Before analysing map A we construct the $(n=1,l=2)$ map by summing 
the powers in vertically adjacent bins and storing this as map B (step 1).
We then search map A using HACR before
summing the power of pixels in horizontally adjacent bins to construct the
$(n=2,l=1)$ map, and overwrite map A (step 2). Repeating this procedure on this new map A, we construct and search all the box
sizes $(n=2^{n_t},l=1)$. Before analysing the $(n=1,l=2)$ map stored as map B we construct
the $(n=1,l=4)$ map and store this as map A (step 3).
Using and overwriting map B, we construct and search all the box sizes $(n=2^{n_t},l=2)$ (step 4).
We repeat this procedure until we have searched all possible box sizes up to the limit imposed by the size of our time-frequency map.

This efficient binning method requires the storage of only two 
time-frequency maps at any given time and reduces the number of floating point operations needed through careful recycling of maps. It is thus very computationally efficient.

We set the HACR thresholds separately for each binned spectrogram and denote them $\eta_{low}^{n,l}$, 
$\eta_{up}^{n,l}$ and $N_{p}^{n,l}$ where the superscripts refer to
the dimensions of the box size we are using. 
A HACR detection occurs if the pixel threshold is exceeded by {\it at least one} cluster in {\it at least one} binned spectrogram.

To characterize the entire search (over all box sizes) we define an overall false alarm probability ($OFAP$). This is defined as {\it the fraction of LISA missions} in which HACR would make {\it at least one false detection} in {\it at least one of the binned time-frequency maps}, in the absence of any signals. Each box size that we use to analyse the data could be allowed to contribute
a different amount to $OFAP$, but to avoid prejudicing our results, we choose to assign an equal false alarm probability to each box size. We call this quantity
the additional false alarm probability ($AFAP$). 
To be clear, $AFAP$ is the probability of a false alarm in a particular box size, i.e., the fraction of LISA missions in which {\it that particular box size} would yield a false detection. The way in which the thresholds are computed ensures that each box size adds $AFAP$ to the overall false alarm rate (hence ``additional''), despite the fact that the binned spectrograms are not all independent. This will be described in Section~\ref{HACRtune}, and ensures that in practice $OFAP=N_{box}*AFAP$.

It is important to note that in the case $N_{p} =1$ then the HACR algorithm 
is equivalent to the Excess Power method described in earlier papers \cite{wengair05,gairwen05}. A 
comparison between these two algorithms will be made in subsequent sections of this paper.  

\subsubsection{Constructing spectrograms}
For our preliminary study we considered a three year LISA mission, and used $3 \times 2^{25}s$ 
of simulated LISA data sampled at $0.125 Hz$ (a cadence of $8s$). To construct the time-frequency map, this data was divided into
$2^{20}s$ ($\sim 2$ week) sections, and an FFT was performed on each section. The resulting time-frequency spectrograms consist of $96$ points in time and $65536$ points
in frequency giving us $N_{box} = 7 \times 17 = 119$ possible
box sizes of the form $n=2^{n_t}, l=2^{n_f}$ where $n_{t}= 0 ... 6$ and $n_{f} = 0 ... 16$. 

A power spectrogram was constructed 
for both LISA low-frequency channels, $h_{I}$ and $h_{II}$ and these were
summed pixel by pixel to produce the time-frequency map searched by the HACR algorithm. In practice, the noise in the two LISA channels was taken to be normal, Gaussian and white but the injected signals were whitened using the theoretical LISA noise
curve described in Section~\ref{LISA}. In this approach, in the absence of a signal the power, $P_{i,j}$, in each pixel of the unbinned spectrogram will be distributed as a $\chi^2$ with 4 degrees of freedom.

The division into $\sim2$ week segments was chosen to facilitate comparison with the Excess Power search \cite{wengair05,gairwen05}, and it is a fairly reasonable choice for EMRIs. The maximum segment length that ensures a source whose frequency is changing at $\rmd f/\rmd t$ does not move by more than one frequency bin over the segment is $1/\sqrt{\rmd f/\rmd t}$. In the extreme mass ratio limit, the leading order post-Newtonian rate of change of frequency is $\rmd f/\rmd t=192/5\,m/M^3\,(M/r)^{11/2}$ for a circular orbit of radius $r$ (in units with $c=G=1$) \cite{peters64}. For the inspiral of a $10M_{\odot}$ object into a $10^6 M_{\odot}$ this gives a maximum segment length of $\sim 1$ day when the orbital radius is $r=10M$. At that radius, the frequency would change by $\sim 10$ bins during one $2$ week time segment. This change is less rapid earlier in the inspiral and more rapid later in the inspiral. If we choose time segments that are too short, the spectrogram will be dominated by short timescale fluctuations in the detector noise, and the frequency bins will be broad, so we lose resolution in the time frequency map. Time segments with length $\sim 1$ week seem like a reasonable compromise. In the future, we plan to experiment with shorter and longer time segments. However, the choice of time segment length should not significantly affect our results thanks to the binning part of the search algorithm.

\subsubsection{Computational cost}
The computational cost of running the HACR search is very low. If we divide the LISA data stream into $M$ time segments, each one containing $N$ time samples, the number of floating point operations required to compute the unbinned spectrogram of the data is roughly $2\,M\,N\,\log N$ using Fast Fourier Transforms. The efficient binning algorithm ensures that only two floating point operations are needed to evaluate the average power for a given pixel in any one of the binned spectrograms (as opposed to $n\times l + 1$ operations if the binned spectrogram was computed directly from the unbinned map). The number of operations required to construct all the binned spectrograms is therefore less than $MN\,\log_2 M \log_2 N$ (less since the average power is not defined for pixels in the last $n-1$ columns and $l-1$ rows when using the $n \times l$ box size). The HACR algorithm first identifies pixels as black, grey or neither ($M\,N$ operations) and then counts the number of pixels in each cluster. For a given cluster, counting the size involves $9$ operations per pixel ($8$ comparisons to see if the 8 possible neighbours are also in the cluster and one addition to increment $N_p$). If HACR has identified $N_c$ clusters, and cluster $c_j$ has $N_p(c_j)$ pixels, this makes $N_c (9N_p(c_j) + 1)$ operations in total, assuming no overlap between the clusters. We do not know in advance how many clusters HACR will identify, nor how many pixels will be in each one, but $N_c < M\,N$ necessarily,  and $N_p < 50000$ since we choose the minimal lower threshold that we use to avoid exceeding this limit (this will be described in the next subsection).

In practice, to run the HACR search with a single set of tuned thresholds on a spectrogram containing a single source, and with LISA and box size parameters as described in the previous subsection, takes about $1$ minute on a $3.5$GHz workstation. If more sources were present, this time would be larger since more clusters would be identified, but $10$ minutes would be an upper limit. This should be compared to the cost of the semi-coherent search which requires $\sim 3$ years on a $50$Tflops cluster \cite{jon04}. It should be noted that noise characterisation and tuning of HACR is more expensive, since it involves using low thresholds (thus increasing the number of HACR clusters identified), and repeating over many noise realisations. However, to complete $1000$ tuning runs using $40$ nodes of a typical computer cluster still takes only a few hours.

\subsection{Search characterisation}
\label{searchchar}
Tuning HACR is a two step process. Firstly, simulated noisy data is analysed in order to identify threshold triplets $\eta^{n,l}_{low}$, $\eta^{n,l}_{up}$ and $N^{n,l}_{p}$ which yield a specified false alarm
probability, $AFAP$, for each box size, $n\times l$. Secondly, a stretch of simulated data containing both noise and a signal is analysed using these threshold triplets and the detection rate (or probability) is measured. For each value of false alarm probability considered we can then choose the threshold triplet that gives the largest detection rate. In this way, we obtain the optimum Receiver Operator Characteristic (ROC) curve for the detection of a particular source.
Throughout the paper, we will use the terms detection {\it rate} and false alarm {\it probability} in order to make a distinction between event candidates caused by a signal or by noise. What we are really computing as the detection {\it rate} is the detection probability, i.e., the fraction of LISA missions in which a particular source would be detected if we had many realisations of LISA. A more relevant observational quantity is the fraction of sources of a given type in the Universe that would be detected in a {\it single} LISA observation. The population of LISA events will have random sky positions, plunge times and plunge frequencies. They therefore sample different parts of the time-frequency spectrogram, which to some extent mimics averaging over noise realisations. The detection rate can thus be taken as a guide to the fraction of sources similar to the given one that would be detected in the LISA mission. A more accurate assessment of the fraction of sources detected requires injection of multiple identical sources simultaneously, but we do not consider that problem here.

To characterize the noise properties of the search we used $10000$ noise realisations and analysed them for twenty choices of $\eta_{low}^{n,l}$ and with the threshold $\eta_{up}^{n,l}$ set as low as is sensibly possible, recording the peak power, $P_{max}$, and size, $N_{c}$, of every cluster detected. With such a list of clusters, it is possible during post-processing to obtain the number of false alarm detections that would be made using any of the twenty lower thresholds, $\eta_{low}$, any value of $\eta_{up} > (\eta_{up}^{n,l})_{min}$ and any value of $N_{p}$. The value of $(\eta^{n,l}_{up})_{min}$ has to be chosen carefully. If it is too low, many clusters will be found in every noise realisation, increasing the computational cost. If it is too high, too few clusters will be identified to give reasonable statistics. We used values of $(\eta^{n,l}_{up})_{min}$ chosen to ensure that a few clusters were found for each box size in each noise realisation.  The lower threshold has to be less than or equal to$(\eta^{n,l}_{up})_{min}$. If it is set too low, large portions of the time-frequency map can be identified as a single cluster. Therefore, we 
choose the minimum value of $\eta_{low}^{n,l}$ to ensure that all clusters are of 
reasonable size, which we define to be less than $50000$ pixels. By examining cluster properties in a few thousand noise realisations, we found suitable empirical choices to be
\begin{eqnarray}
\alpha^{n,l} & = & 4 + \frac{10 \sqrt{2}}{(n l)^{5/9} } \\
(\eta_{low}^{n,l})_{min} & = & 4 + 4 \sqrt{ \frac{10}{50000 + n l} } \\
(\eta_{up}^{n,l})_{min}  & = & \max[ \alpha^{n,l} , (\eta_{low}^{n,l})_{min} ]. 
\end{eqnarray}
We note that for large box sizes, $\alpha^{n,l} < (\eta^{n,l}_{low})_{min}$ and so we set $\eta_{up}^{n,l} = \eta_{low}^{n,l}$. Above this point, we no longer ensure that at least one cluster is found for each box size, as this is inconsistent with the more important requirement that no cluster exceeds $50000$ pixels. We emphasise that our search is robust to the somewhat arbitrary choice of these minimal values. For box sizes where $(\eta_{low})_{min} < (\eta_{up})_{min}$, we use 20 values of $\eta_{low}$ spaced evenly between $(\eta_{low})_{min}$ and $(\eta_{up})_{min}$. Where $(\eta_{low})_{min}=(\eta_{up})_{min}$, we use only one lower threshold $\eta_{low}=(\eta_{up})_{min}$. This comparatively small number of lower threshold choices is sufficient to find the maximum detection rate thanks to the degeneracy between $N_p$ and $\eta_{low}$ mentioned earlier.

\subsection{Post-processing} 
\begin{figure}[!hb]
\begin{center}
\includegraphics[height=5in,angle=-90]{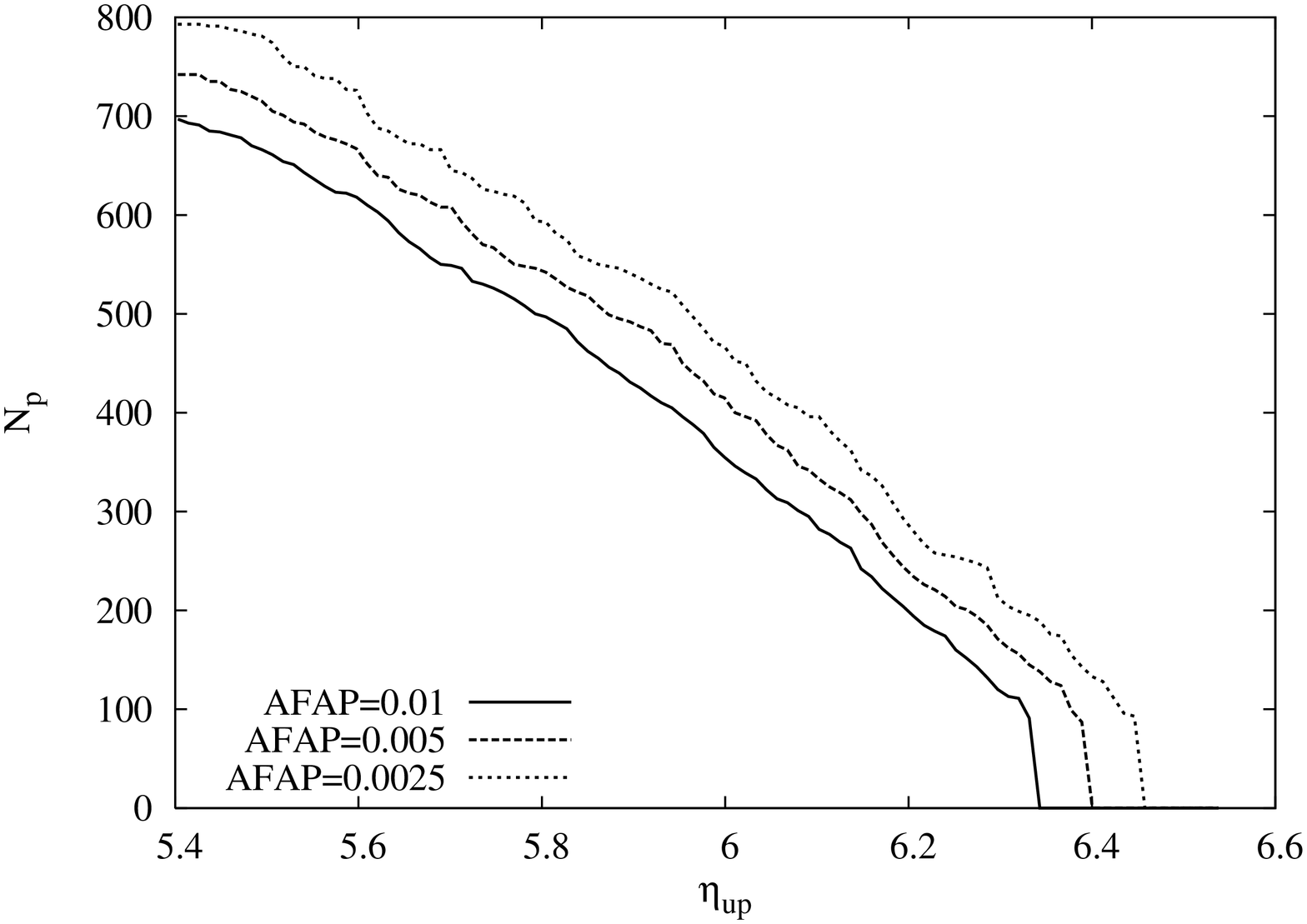}
\includegraphics[height=5in,angle=-90]{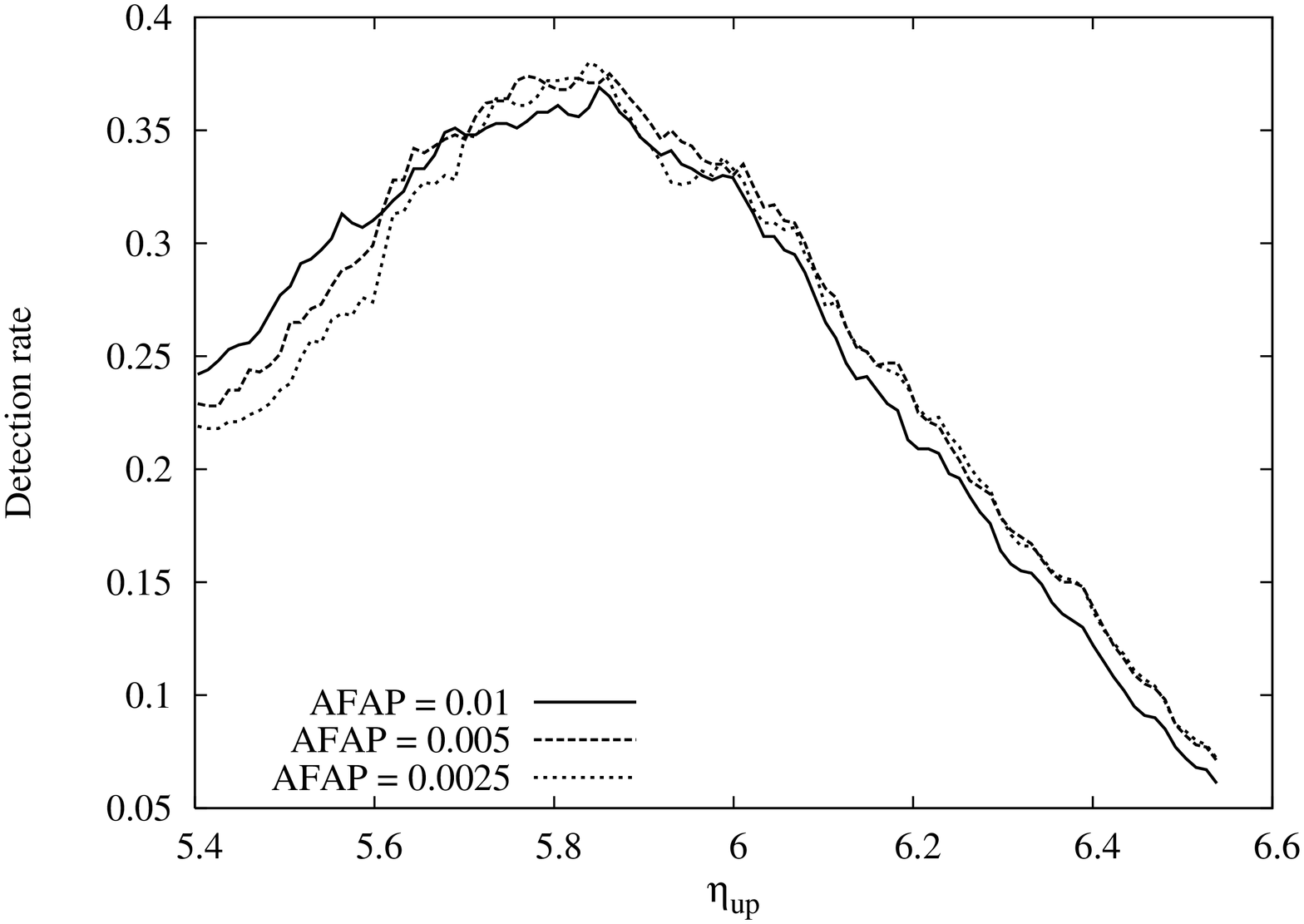}
\end{center}
\caption{Top panel: Contours of constant (additional) false alarm probability for the box size $n=1,l=64$ and one particular lower threshold value. Pairs of thresholds $\eta_{up}$ and $N_{p}$ which lie on a contour yield the same additional false alarm probability. Bottom panel: Rate of signal detection plotted against choice of threshold, again for fixed lower threshold. Each point on the x-axis represents a set of thresholds which yield a particular value of $AFAP$. By choosing the threshold set which yields the largest value of detection rate, plotted on the y-axis, we can maximise the rate of signal detection for a given false alarm probability.
}
\label{threshrel}
\end{figure}

For each box size and each lower threshold value we can consider values of $\eta_{up}^{n,l}$ between 
$(\eta_{up}^{n,l})_{min}$ and the maximum power measured 
$(\eta_{up}^{n,l})_{max}$, and construct a list of all clusters with 
peak power $P_{max} > \eta_{up}^{n,l}$, ordered by increasing number of 
pixels $N_{c}$. If the false alarm probability in a given box size, $AFAP$, has been chosen, we expect to see $M\times AFAP$  false alarms in $M$ noise realisations. By looking at the list of clusters, we can identify a value of the threshold $N_{p}^{n,l}$ with each pair of values for $\eta_{up}^{n,l}$ and $\eta^{n,l}_{low}$ that would give the correct number of false alarms in the realisations considered. If HACR was used to analyze pure noise with those three thresholds and only that one box size $(n,l)$, it would yield a detection rate $AFAP$. A typical relationship between 
$\eta_{up}^{n,l}$ and $N_{p}^{n,l}$ for a fixed choice of $\eta_{low}^{n,l}$ is shown in the left panel of Figure~\ref{threshrel}. This was generated for a box size $n=1$, $l=64$, the $6$th lower threshold value of the $20$ examined, and for three choices of $AFAP = 0.01, 0.005$ and $0.0025$.

To determine which combination of thresholds is optimal, we subsequently analyse spectrograms containing both noise and an injected signal. As mentioned earlier, since we are using white noise to generate the noise realisations, the signal is first whitened using the noise model described in Section~\ref{LISA} before injection. For each box size we may then select the triplet of thresholds which yields the largest detection rate. The right panel of Figure~\ref{threshrel} shows detection rate plotted against upper threshold value for EMRI source ``A'' at a distance of 2Gpc using the box size $n=1$, $l=64$ with $AFAP = 0.01, 0.005$ and $0.0025$ and for a fixed lower threshold value (the $6$th of the $20$ values used). Although only the $\eta_{up}$ threshold value is shown a corresponding value of $N_{p}$ is inferred, determined by the assigned $AFAP$. This stage of the analysis will be discussed further in the next section.

The full search uses multiple box sizes, searched in a particular order. We want the thresholds in a given box size to contribute an {\it additional} false alarm rate of $AFAP$. When determining the threshold triplets we therefore need to ignore realisations in which false alarms have already been found. The procedure above is thus slightly modified when considering more than one box size. If we are using $M$ noise realisations to determine the thresholds, each box size should give $M\times AFAP$ false alarms. The necessary threshold triplet can be determined for the first box size as described above. It is then possible to identify the realisations in which the false alarms were found for the first box size. This set of realisations will be somewhat different for each of the triplets of thresholds that give the desired $AFAP$. So, in practice we must do this in conjunction with the source tuning described in the next section. This allows us to identify an {\it optimal threshold triplet} and we can find the noise realisations in which {\it that} threshold triplet gave false alarms. We then repeat the procedure described above, but now considering only clusters identified in the {\it remaining} realisations. This process is repeated for each box size in turn, ignoring in each subsequent box size any realisations in which false alarms have been identified in earlier box sizes. This means that the order in which the different box sizes are searched affects the thresholds. However, our results show that it does not matter in which order the box sizes are searched, provided the order is the same for tuning and the actual search. This will be discussed further in Section~\ref{compPerformance}. 

\section{Performance of HACR in EMRI detection}
\label{EMRIperformance}

\subsection{Tuning HACR for a single specific source}
\label{HACRtune}
The fact that HACR has three thresholds allows the search to be tuned to
optimally detect a specific source at a specific distance. For a given choice
of false alarm probability, $AFAP$, we can choose the triplet of thresholds for
each box size $\eta_{low}^{n,l}$, $\eta_{up}^{n,l}$ and $N_{p}^{n,l}$
that maximises the detection rate.
For this optimal threshold triplet, a Receiver Operator Characteristic (ROC) curve can be plotted for the HACR search tuned for that source. The ROC curve shows the detection
rate as a function of the overall false alarm probability, $OFAP$, of the
search using all box sizes.

In practice, the ROC is determined by generating a sequence of noise
realisations, injecting the whitened signal into each one, and then
constructing and searching the binned spectrograms.
A detection is defined as any realisation in which all thresholds are exceeded
in at least one box size.
The box sizes are searched in the order they were constructed (see
Figure~\ref{binning}). As discussed in the previous
section, if a detection has been made for one box size, we want to ignore that
realisation when we search with subsequent box sizes. This ensures that we
always choose the threshold triplet for a box size that provides the
maximum number of {\em additional} detections. In practice, we achieve this
goal using the following algorithm
\begin{itemize}
\item Search all realisations using the first box size, for threshold triplets (typically $\sim 100$ upper thresholds and $20$ lower thresholds) that all yield the assigned $AFAP$.
\item Choose the threshold triplet that yields the highest detection rate. Identify every realisation in which {\it this optimal threshold triplet} gives a detection.
\item Move onto the second box size and repeat this procedure, but only
search realisations in which the optimal threshold triplet for the first
box size did {\it not} yield any detections. 
\item Repeat for all other box sizes in order.
\end{itemize}
Once the optimal threshold triplets have been determined in this way, the detection rate must be measured by using these optimal thresholds to search a {\it separate} set of signal injections, to avoid biasing the rates. We experimented with using different numbers of injections and concluded that using 1000 signal injections to determine the thresholds and another 1000 signal injections to measure the rate gave reliable answers. We estimate the error in the resulting ROC curve due to noise fluctuations to be less than $3\%$. All the results in this paper are computed in this way. To characterize the noise, we use the same set of 10000 pure noise realisations in all calculations.

\begin{figure}[!hb]
\begin{center}
\includegraphics[height=5in,angle=-90]{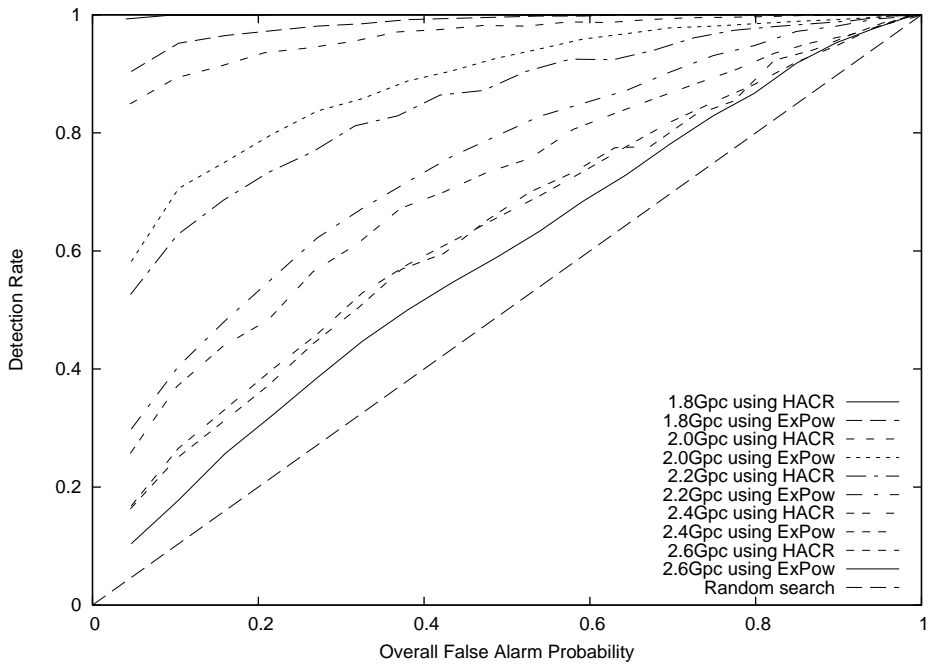}
\end{center}
\caption{Receiver operator characteristic (ROC) curves for detection of an EMRI (source ``A'')
at a range of distances from Earth. For each distance we show ROCs for HACR and the Excess Power search. 
As expected HACR's performance either matches or exceeds that of the Excess Power search.
To aid interpretation of the ROC curve plots in this paper, the ordering of the labels in the legend reflects the
performance of the corresponding ROC curves, i.e. the second label from the top corresponds to the ROC curve
with the second best performance.}
\label{ExPowHACRComparison}
\end{figure}

In Figure~\ref{ExPowHACRComparison} we show the ROC curves for detection of
source ``A'' at a range of distances using HACR.
The random search line on this figure represents a search for which the
detection rate and false alarm rate are equal.
This is the ``random limit'' since it is equivalent to tossing a coin and
saying that if it is heads the data stream
contains a signal and if it is tails it does not.
A search that yields a ROC curve equal to this random line is essentially
insensitive to signals.
In Figure~\ref{ExPowHACRComparison}, we see that the source has a $100\%$
detection rate for all $OFAP$'s out to a distance of $\sim1.8$Gpc.
An overall false alarm probability of 10\% is probably quite a conservative
value,
since this is the probability that {\it in a given LISA mission} the entire
HACR search would yield just a single false alarm.
At a distance of 2Gpc, with the overall false alarm probability set to $10\%$,
HACR achieves a detection rate of $\sim 90\%$.
As the distance increases further, the detection rate further degrades, and the
source becomes undetectable at a distance of $\sim 3$Gpc.
The rate of EMRI events is somewhat uncertain, but the range for a
$10M_{\odot}$ black hole falling into a $10^6M_{\odot}$ black hole
is between $10^{-7}$ and $10^{-5}$ events per Milky Way equivalent galaxy per
year \cite{freitag01,jon04}.
Using the same extrapolation as in \cite{jon04}, this gives $0.1 - 10$ events
Gpc$^{-3}$ yr$^{-1}$.
Assuming a 3 year LISA mission, and that the detection rates quoted here are a
good approximation to the fraction of
EMRI events that LISA would detect in a single realisation of the mission,
these rates translate to a detection of
$\sim 15$--$1500$ events using this method (using a Euclidean volume-distance
relation). We note, however, that at
the high end of this range, source confusion will be a significant problem and
it has been ignored in the current work.

\subsubsection{Comparing the performance of HACR and the Excess Power method}
\label{compPerformance}
In Figure~\ref{ExPowHACRComparison} we also show ROC curves for using the Excess Power search to detect source ``A'' at a range of distances. Since HACR effectively performs the
Excess Power search when $N_{p}^{n,l} = 1$ we expect that HACR will always do
at least as well as the Excess Power search.
Due to the extra levels of tuning allowed by the HACR algorithm we find that it
can obtain a somewhat higher detection rate for a given false alarm
probability. 
The increase is in the range of $5-20\%$ for an $OFAP$ of $10\%$, but this translates to a significantly enhanced event rate.
With the source at a distance of 1.8Gpc both methods achieve very high detection rates; both have detection rates $>95\%$ 
with an $OFAP$ of $10\%$.
At intermediate distances (e.g. $\sim$2.2Gpc) HACR outperforms Excess Power considerably, but once the source is
at 2.6Gpc, there is very little difference in the performance of the two
searches.
However, as illustrated in Figure~\ref{threshrel}, the optimal HACR pixel
threshold tends to be significantly greater than $1$.
Thus, HACR identifies clusters containing significant numbers of pixels, while the
Excess Power search at the first stage identifies only individual pixels. Parameter extraction from the Excess Power method requires an additional track
identification stage. Such algorithms are currently being investigated \cite{wen06}, but HACR is
more efficient, combining both stages into one.
The information contained in the structure of HACR clusters should allow
parameter estimation which can be used as input for later stages in a
hierarchical search. This will be discussed in more detail in Section~\ref{paramest}.

As mentioned earlier, the fact that realisations in which detections are made
are omitted for the search of subsequent box sizes
treats the earlier box sizes preferentially.
We repeated the ROC calculation with the box search order randomized in various ways. It was clear
from this that the overall search performance was independent of the box size search order.
We recommend using the order given by the efficient binning algorithm described
earlier because of the computational savings.

\subsection{Targeted searches}
\label{targsearch}
\begin{figure}[!hb]
\begin{center}
\includegraphics[height=5in,angle=-90]{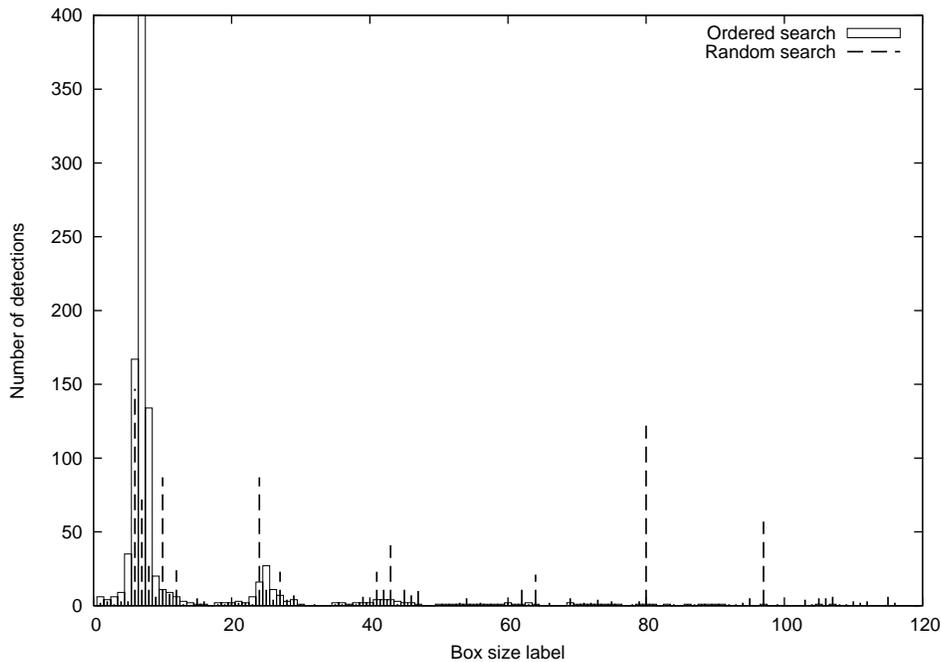}
\end{center}
\caption{Number of detections as a function of box size when searching 1000 realisations of source ``A'' at 2Gpc. Results are shown when using the ordered search, and when the box size search order is randomized. The x-axis is the box size label, which corresponds to the order in which the boxes are analysed in the {\it ordered} search.}
\label{RateVsBinLab}
\end{figure}

In Figure~\ref{RateVsBinLab} we show how the detection rate depends on the box
size.
This figure shows the number of detections made for each box size over the 1000
realisations used for
determination of the ROC curve for source ``A'' at 2Gpc.
It is clear that there is not only one single box size that makes detections, but several box sizes are important.
This is because random noise fluctuations will sometimes make one box size
better than another.
However, it is also clear that many of the box sizes do not make any detections
and are apparently not very useful for the
detection of this particular source.
This is partially due to the box size search order.
Figure~\ref{RateVsBinLab} also shows the detection rate as a function of the
box size label when the search order was randomized.
Although the distribution is qualitatively similar, the box sizes that make
the detections are different in this case. It is clear that there are several box sizes that are equally good at detecting this source (these have approximately the same dimension in frequency, but
different dimensions in time).
Whichever of these equivalent box sizes is used first will make the detection,
although the overall number of detections and ROC performance is independent of
the search order.

Given that we have specified thresholds so that each box size contributes
equally to the overall false alarm probability we might expect the search to perform better if
we restrict it to use only those few box sizes responsible for most of the
detections of the injected signal. By eliminating box sizes that make few
detections, we expect to reduce the overall false alarm probability while
keeping
the overall detection rate approximately constant, thereby improving the
overall ROC performance. This can
be investigated by re-analysing the data using only a small subset ($20$)
of the 119 box sizes originally considered, that were responsible for the most
detections of EMRI source ``A''. Having performed the search using only $20$ box sizes,
we can eliminate the box size which has the worst performance (i.e., the least
number of detections) {\it in the $20$ box search} and then repeat the
search with the remaining $19$ box sizes. This process can be repeated,
eliminating one box size each time, until only one box size remains. The
box size that contributes the fewest detections depends to a limited extent
on the (additional) false alarm probability assigned to each box size.
We used the additional false alarm probability that gave an overall search
false alarm probability of $\sim 10\%$ since, as argued earlier, this would
be a reasonable value to use in the final LISA search.

The results of this targeting procedure are summarized in Table~\ref{TargSearchROC}.
When the number of box sizes is reduced from $119$ to $20$, the ROC performance
does improve as the overall FAP reduces,
while the detection rate remains largely unchanged. This improvement
is of the order of $5\%$ in detection rate.
As the number of box sizes used is reduced further, the ROC performance remains
roughly constant until only $4$ box sizes are being used. Using fewer than $4$ box sizes leads to performance that degrades and is always worse than the full search.
This is in keeping with the understanding that several box sizes are needed for
efficient detection of a source due to the effect of noise fluctuations.
We also computed results for the Excess Power search (full and targeted), and these are also summarized in the table. The trend as box sizes are removed is the same, but the HACR search always outperforms the Excess Power search

We conclude that it is possible to improve the performance of the search for a
specific source by targeting to fewer box sizes.
However, the improvement is not hugely significant.
This is consistent with what was found for the Excess Power search
\cite{gairwen05}.
Since the box sizes that are efficient for the detection of one particular
source will almost certainly not be the same as those
that are efficient for other sources, the best approach is to include all the
box sizes in the search.
However, since there are certain box shapes that are good for detecting certain
types of source, the box size for which a given
detection is made provides a diagnostic of the source system.

\begin{table}[!hb] 
\begin{tabular}{c|c|c|c|c}
Search&\multicolumn{4}{c}{Detection rate at}\\ \hline
&OFAP=$5\%$&OFAP=$10\%$&OFAP=$30\%$&OFAP=$60\%$ \\ \hline
HACR, All bins& 84.9\%&89.3\%&95.5\%&98.7\% \\ \hline
HACR, 20 bins&90.2\%&92.9\%&98.2\%&99.7\% \\ \hline
HACR, 10 bins&90.5\%&93.4\%&98.4\%&99.6\% \\ \hline
HACR, 7 bins&92.0\%&94.7\%&98.4\%&99.4\% \\ \hline
HACR, 4 bins&92.7\%&95.0\%&98.5\%&99.4\% \\ \hline
HACR, 1 bin&81.7\%&87.5\%&95.2\%&99.0\% \\ \hline
Excess Power, All bins&63.8\%&71.5\%&87.1\%&95.4\% \\ \hline
Excess Power, 10 bins&72.6\%&81.4\%&94.0\%&98.2\% \\ \hline
Excess Power, 7 bins&66.0\%&76.0\%&91.0\%&98.1\% \\ \hline
Excess Power, 4 bins&68.7\%&78.5\%&91.3\%&98.4\% \\ \hline
Excess Power, 1 bin&47.8\%&59.1\%&79.7\%&93.8\% \\ \hline
\end{tabular}
\caption{Detection rates for various overall false alarm probabilities when using the HACR or Excess Power searches with a restricted number of box sizes. } 
\label{TargSearchROC}
\end{table}

\subsection{Detection of other EMRI sources}
\begin{figure}[!hb] \begin{center}
\includegraphics[height=5in,angle=-90]{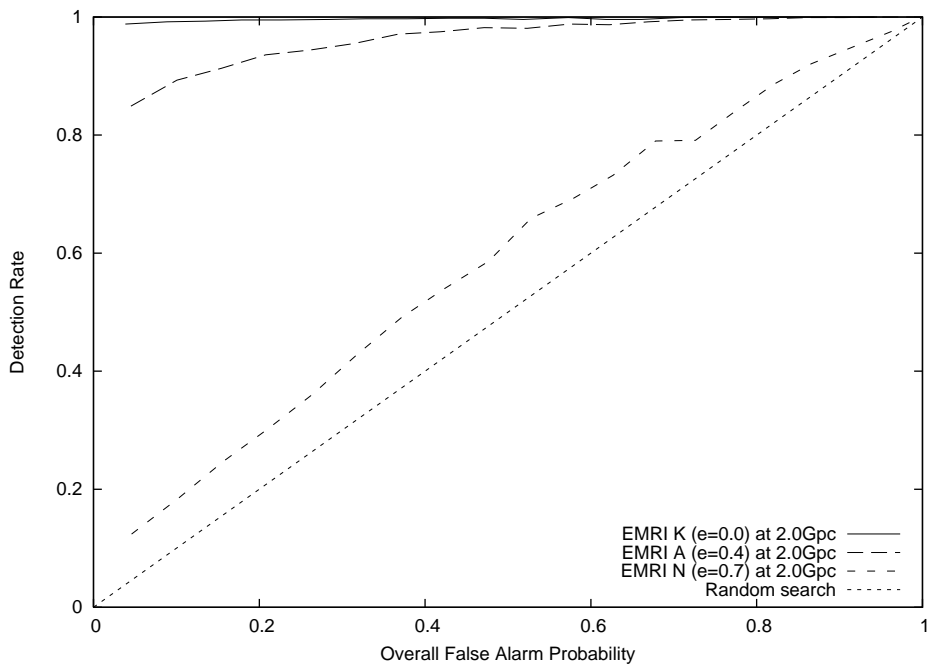}
\end{center} 
\caption{ROC curves for detection of EMRI sources ``A'', ``K'' and ``N'' at a distance of 2Gpc using HACR.
These sources all have the same parameters except for their eccentricity. } 
\label{SigAKN_2GpcROC}
\end{figure}

The results described in the preceding sections have focused on the detection
of one particular EMRI, source ``A''.
We have also explored the performance of HACR in detecting some of the other EMRI sources used for the investigation of the Excess Power search \cite{gairwen05}.
Specifically we used the sources ``K'' and ``N'', which have the same
parameters as source ``A'' except for eccentricity.
The source ``K'' is initially circular, while source ``N'' has eccentricity
of $0.7$, compared to $e=0.4$ for source ``A''.
We placed these sources at a range of distances between 1.8Gpc and 2.6Gpc, and
injected them into noise realisations.
We were thus able to determine ROC curves for detection of these sources via the method described in Section~\ref{HACRtune}.
In Figure~\ref{SigAKN_2GpcROC} we compare the ROC curves for detection of these
sources with HACR when they are at a distance of 2Gpc.
We see that our ability to detect a system at a given distance is better for
binaries in circular orbits (source ``K'')
than for systems with eccentric orbits (sources ``A'' and ``N''). This is
consistent with what was found for the
Excess Power search in \cite{gairwen05}.
The predominant effect of orbital eccentricity is to split the GW radiation
power into multiple harmonics. As the
eccentricity increases, the frequencies of these harmonics become increasingly
separated. As a consequence, a given
box in the time-frequency map contains a smaller ratio of signal power to noise
power. The detectability of EMRI
sources therefore decreases as the eccentricity is increased.

The performance of the HACR search for other sources considered in
\cite{gairwen05} is similar in general. The overall
detectability follows the same pattern as the Excess Power search.
HACR has a slightly greater detection rate than Excess Power when the source is
nearby, but as the source is put further away,
the performance of HACR and Excess Power become comparable before the random
limit is reached.
However, in all cases, the HACR detection is made with a smaller upper threshold ($\eta_{up}^{n,l}$) than Excess Power, compensated by a
larger pixel threshold ($N_p^{n,l}$).
Thus, HACR detections identify clusters with significant numbers of pixels, the
properties of which will be
invaluable for subsequent parameter estimation.
This will be discussed in Section~\ref{paramest} and will also be the subject
of a follow-up paper that is currently in preparation.

\subsection{Tuning HACR for multiple sources}
\label{multEMRItune}
In the preceding sections, we have focused on detection of a single source at a
fixed distance.
However, in the actual analysis of LISA data, we will not know a priori what
sources will be in the data stream, and so the
HACR thresholds need to be tuned as generally as possible.
Even in the case of a single EMRI source, the optimal threshold combination
depends to some extent on the distance at which the source is placed.
This is in contrast to the Excess Power search, where there is only one
threshold that is uniquely determined by the choice of false alarm
probability.
There are two possible approaches to constructing a general HACR search --- 1)
have several separate HACR searches, targeting different
sources and using different sets of thresholds or 2) have a single HACR search
with a set of thresholds chosen to be sensitive to as many
LISA sources as possible.
To date, we have focussed on the latter approach, since our results have shown
that it is possible to do almost as well with a single set
of ``generic'' thresholds as with source specific thresholds.

As a first step, we took the thresholds designed to optimally detect source ``A'' at 2Gpc and used those thresholds to search for sources ``K'' and ``N''. We found that there was some degradation of performance, but that this was negligible. At an $OFAP$ of $10\%$, the detection rate for source ``K'' changed from $99.3\%$ to $99.7\%$, and that of source ``N'' changed from $18.4\%$ to $17.9\%$. This is a promising result and suggests that certain threshold combinations do well at detecting all the EMRI events. It is also possible to tune the thresholds to be generally sensitive to many different sources. This is not really necessary for the case of EMRI detection, but we will describe the procedure here as it will be needed when other types of source are included in the search. 

We want to tune the search to maximize the total LISA event rate
(i.e., the number of events observed).
If we knew in advance which sources would be present in the LISA data, we could
tune the search by considering multiple noise
realisations with that family of sources injected and choosing the threshold
combination that gives the maximum total detection
rate for given $OFAP$. Since we do not know what the actual sources in
the LISA data will be, we can instead tune the thresholds
to be as sensitive as possible to a single event of unknown type, using
prior knowledge to weight the relative likelihood of
different types of event.
This procedure ignores issues of source confusion, but should ensure that the
loudest events are detected, no matter of what type or
at what distance they might be.

In practice, tuning for multiple sources is done as follows:
\begin{itemize}
\item Generate realisations of noise with injected signals for each of the sources we want to include in the tuning.
\item For the first box size, determine the rate of detections, $R_s({\bf t_i})$, of each of the signals when using HACR with each threshold triplet, ${\bf t_i}$, that yields a pre-chosen $AFAP$.
\item Construct a sum over these rates for each threshold triplet, $\sum w_s R_s({\bf t_i})$, using an appropriate weighting factor, $w_s$, for each source.
\item Choose the threshold triplet that maximizes this weighted sum. For each signal, identify the realisations in which that optimal threshold triplet gave a detection.
\item Move onto the next box size, but for each signal search only realisations in which the optimal thresholds for the previous box size(s) did not yield any detections.
\item Repeat for all box sizes.
\end{itemize}
One question is what to use for the weighting factors.
If we knew that only one type of source existed in the Universe, but it was
equally likely to be at any point in space,
we want a volume weighted average.
This is done by taking our set of sources to be a single given source placed at a
sequence of distances, $d_i$.
The source at distance $d_i$ can then be regarded to be representative of all
sources in the range $d_{i-1} < d < d_i$, and
should be weighted by the (Euclidean) volume of space in that range, $w_{i}
\propto 4\pi(d_i^3 - d_{i-1}^3)/3$.
We carried out this procedure using source ``A'' at distances of $1.8$Gpc,
$2.0$Gpc, ..., $2.6$Gpc, with
weightings $1.8^3 = 5.832$, $2.0^3 - 1.8^3 = 2.168$, $2.2^3 - 2.0^3 = 2.648$
... $2.6^3 - 2.4^3 = 3.752$
(we have neglected common factors of $4 \pi / 3$). We took the closest source to be at $1.8$Gpc since up to that distance, the detection rate is always $100\%$.
This appears to give artificial weight to the $1.8$Gpc source, but in practice this does not happen since virtually every threshold combination gives a $100\%$ detection rate for that source, and
the variation in rate is determined primarily by the other injections.
We used distance weighted thresholds to search for source ``A'' at various distances. The thresholds did change to some extent, but these changes were small since the optimal thresholds are almost independent of distance, and the overall ROC performance was largely unaffected. We deduce that it is possible to detect a given EMRI source at any distance with a single set of thresholds.

LISA will see more than one type of source, and we can fold in prior information
about the relative abundance of different
events by adjusting the weighting factors.
We repeated the above, tuning for sources ``A'', ``K'' and ``N'' at a single distance
of $2$Gpc, and given equal weighting.
In that case too, we found that the ROC performance was not significantly
changed when tuning for these multiple sources.
We also tuned for all three sources, placed at all the
distances, $1.8$Gpc, ..., $2.6$Gpc, with the volume
weightings listed previously.
Once again, the ROC performance was not significantly altered.
Thus, there is a single set of HACR thresholds that can detect all three EMRI
sources at any distance.

These results may not be truly generic, since the three EMRI sources are quite
similar, differing only in eccentricity.
It is therefore perhaps unsurprising that a single set of thresholds can detect
all three sources almost optimally.
However, we will see in Section~\ref{multtune} that this result carries over to the case when the sources have quite different characteristics. This is not totally surprising, since we know that HACR includes the Excess Power search as the pixel threshold $N_p=1$ limit. The Excess Power search thresholds are independent of the tuning source at fixed assigned FAP. Thus, a HACR search tuned for a collection of sources can do no worse than the Excess Power search for each of those sources.
Since the HACR search does not seem to hugely outperform the Excess Power
search, we would not anticipate
that this combined tuning procedure would lead to a serious degradation of
performance even when considering very different classes of source.

\section{Performance of HACR in detection of other LISA sources}
\label{otherperformance}
We have shown that HACR may be successfully tuned in order to detect multiple EMRI sources with different parameters. In this section we investigate HACR's ability to detect other classes of signals, specifically white dwarf (WD) binaries
and supermassive black hole (SMBH) binary mergers. We expect these other classes of signal to have quite different structure in a time-frequency map. A typical EMRI signal consists of several frequency components (due to the eccentricity of the orbit), which ``chirp'' slowly over the course of the observation, i.e., the frequency and amplitude increase. By contrast, the GW emission from a WD binary is essentially monochromatic. A SMBH binary inspiral also gives a chirping signal, but the chirp occurs much more quickly than the EMRI due to the increased mass ratio, so it will be characterised by a signal that is broader in frequency. This difference in structure allows HACR to be tuned for all three types of source simultaneously.

\subsection{A typical SMBH binary source}
As a preliminary investigation, we repeated the tuning procedure described earlier, injecting a typical SMBH binary inspiral and a typical WD binary at various distances. The SMBH binary waveform represented the inspiral of two $10^6 M_{\odot}$ non-spinning black holes, placed at a random sky position, and with merger occurring $\sim3$ weeks before the end of the observation. As mentioned in Section~\ref{sourcemodel}, our SMBH injections use the waveform model given in \cite{curt98}. This is a restricted post-Newtonian waveform accurate to 1.5PN. More accurate waveforms are available in the literature, with post-Newtonian corrections up to 3.5PN. However, the simple model captures the main features of a SMBH merger signal and is accurate enough for the more qualitative nature of this preliminary study. The quoted masses are the intrinsic masses of the black holes, i.e., not redshifted. When the source was placed at higher redshift there are therefore two effects --- an increase in the luminosity distance to the source, and a redshifting effect --- which pushes the signal into the less sensitive part of the LISA noise curve. In Figure~\ref{SMBHWDAllDist} we show the ROC curves for detection of this SMBH binary source at a range of redshifts. At each redshift the optimal thresholds were chosen using the tuning method described in Section~\ref{HACRtune}. We find that SMBH binary sources at redshifts $z \leq 3$ are detected with almost perfect efficiency using HACR, but we stop being able to resolve signals for redshifts $z > 3.5$. This is primarily because the (matched filtering) SNR of the source decreases significantly due to the redshifting effect mentioned above.

\subsection{A typical WD binary source}
\begin{figure}[!hb]
\begin{center}
\includegraphics[width=3.5in,angle=-90]{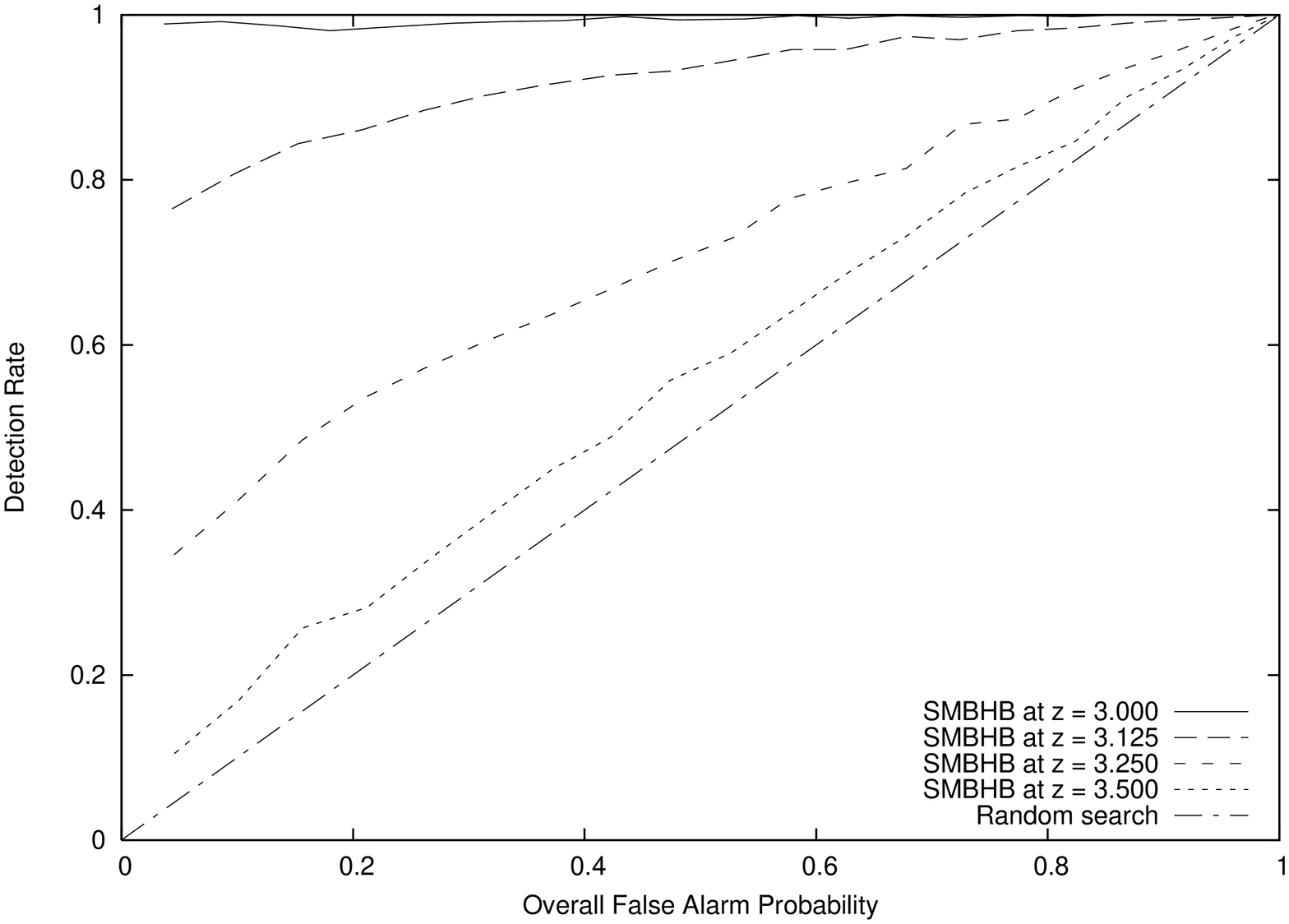}
\includegraphics[width=3.5in,angle=-90]{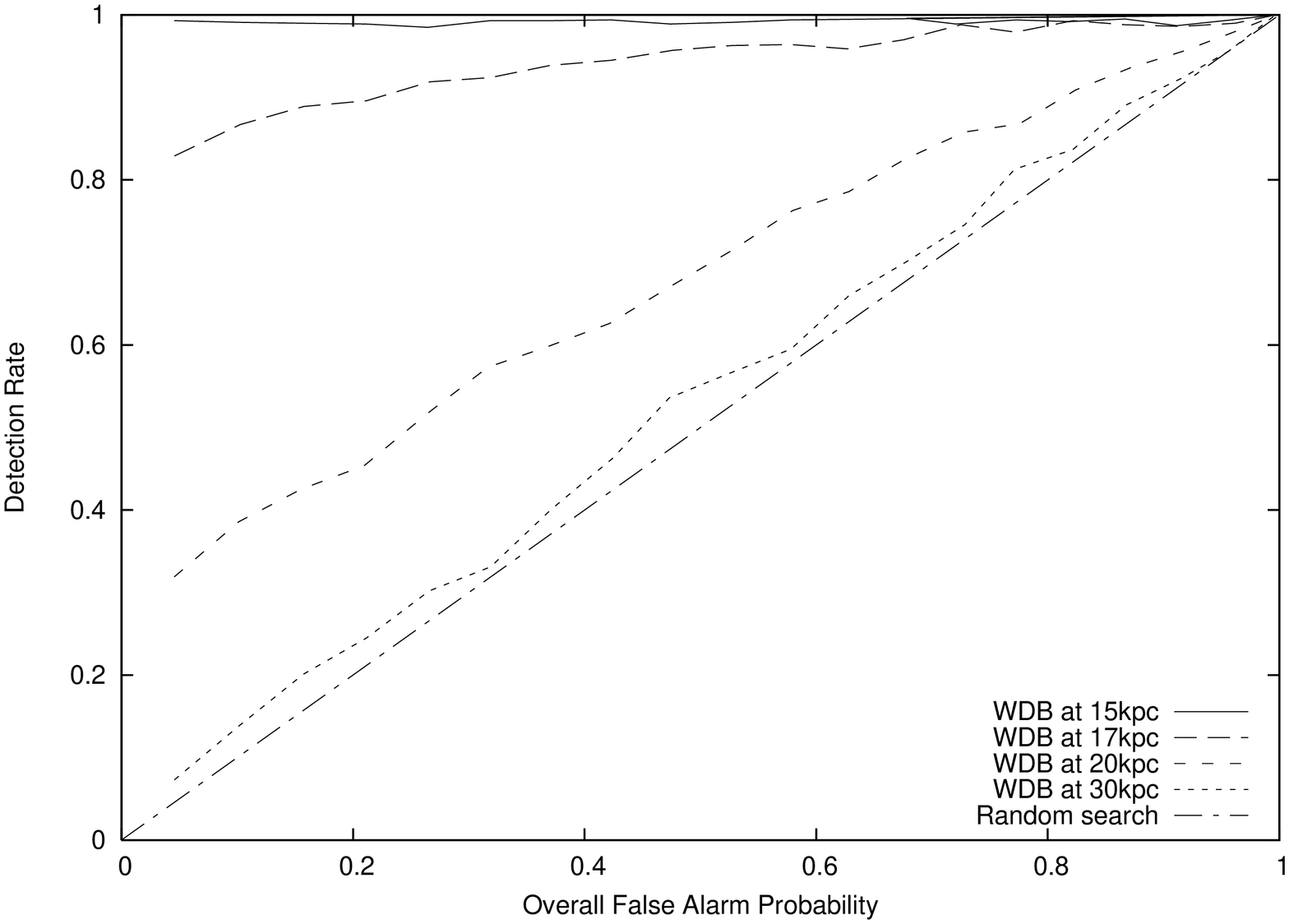}
\end{center}
\caption{ROC curves for detection of a SMBH binary merger (left panel) and a WD binary (right panel) at various distances. The optimal thresholds for each distance were chosen using the
tuning method described in section \ref{HACRtune}}
\label{SMBHWDAllDist}
\end{figure}

The ``typical'' WD binary was chosen to have the parameters of RXJ0806.3+1527 (as quoted in \cite{stroeer06}), except for distance and sky position. The latter was chosen randomly, but this choice, and the noise model used meant the SNR of this source at a distance of $1$kpc was approximately a factor of $3$ greater than that quoted in \cite{stroeer06}. This should be born in mind when considering the distances quoted in the following discussion. In Figure~\ref{SMBHWDAllDist} we show the ROC curve for this WD source, injected at various distances. At distances $ \leq 15$kpc, we obtain near perfect detection using HACR. The sensitivity falls off rapidly for greater distances and the source becomes undetectable at greater than $\sim 20$kpc. Even allowing for the SNR discrepancy mentioned above, this source would be detectable at $\sim6$--$7$kpc, so almost at the distance of the galactic center. Since this particular source is estimated to be at a distance of $300$--$1000$pc, it would be detectable via this method. We would expect to detect other similar white dwarfs at distances of $1$--$10$kpc depending on the source parameters. This does not allow for source confusion, as we have only injected single sources into the data stream, but the conclusion for RXJ0806 should be robust, since it radiates at $\sim 6$mHz, which is in the regime where WD binaries are well separated in frequency (this can be seen in the results of population synthesis models described in \cite{nelemans01} and is reflected in the LISA noise curve \erf{LISAsh} in which the contribution from WD binaries, accounting for resolvability of sources, is below the instrumental noise at $6$mHz).

\begin{figure}[!hb]
\begin{center}
\includegraphics[height=5in,angle=-90]{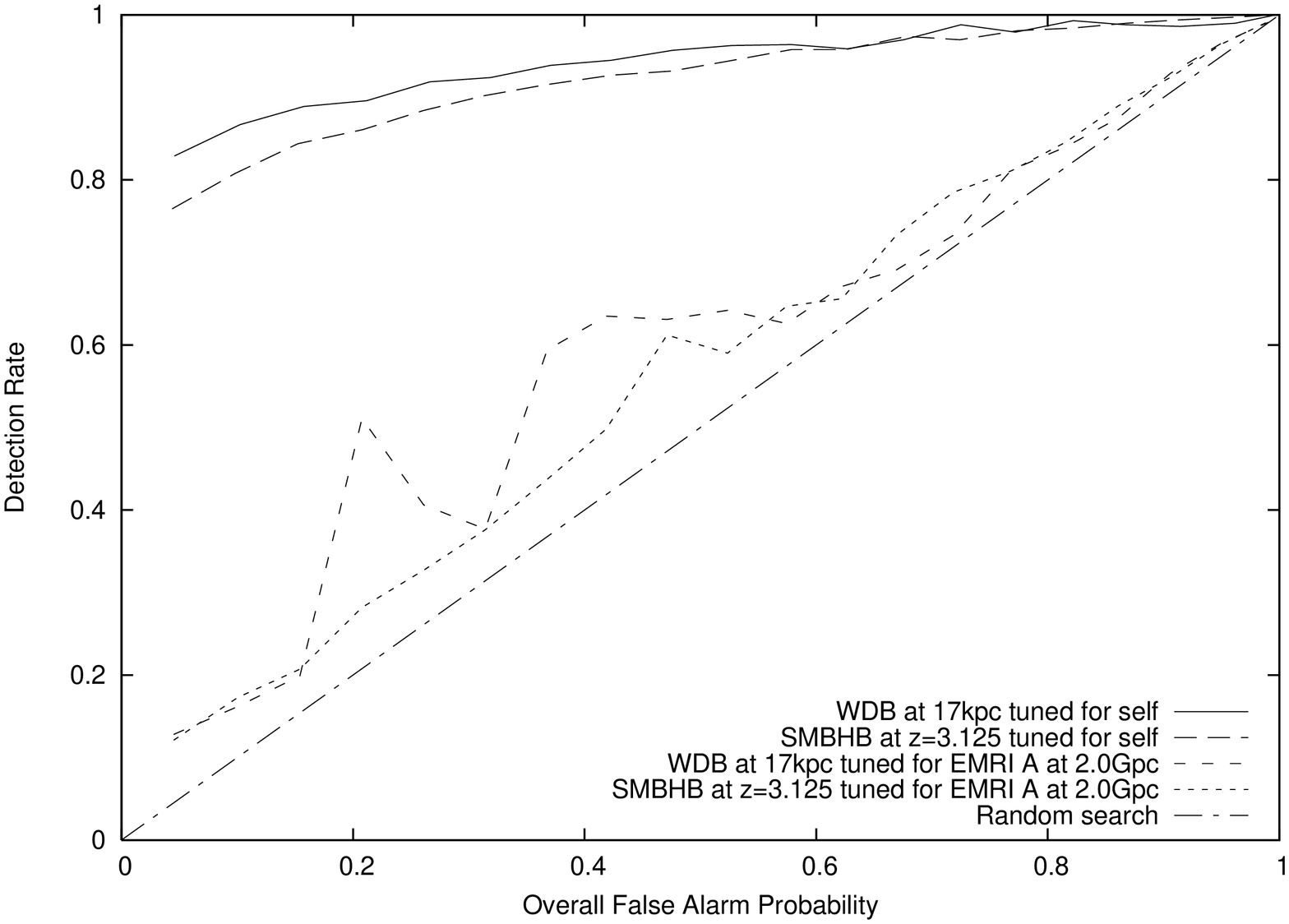}
\end{center}
\caption{ROC curves for detection of the SMBH and WD binary sources using thresholds tuned for EMRI source ``A'' at a distance of 2Gpc. For comparison, the ROC performance when the search is tuned for the source in question is also shown.}
\label{TestSigA2GpcThresh}
\end{figure}

In the preceding plots, the HACR thresholds have been tuned to detect the source in question, at a particular distance. If instead we imagined that we would use only one set of thresholds, tuned for EMRI source ``A'' at a distance of 2Gpc, then the ROC performance for detection of the SMBH binary and WD binary events is significantly degraded. This is shown in Figure~\ref{TestSigA2GpcThresh}, which compares the ROC curve for detection of the SMBH binary at redshift $z = 3.125$ and the WD binary at a distance of $17$kpc when the EMRI thresholds are used, versus the result when the source specific tuned thresholds are used. We chose distances of $z=3.125$ and $17$kpc since in that case the sources are loud, but have less than a $100\%$ detection rate, so we will be able to see ROC variations. Figure~\ref{TestSigA2GpcThresh} shows that using the EMRI thresholds to detect other sources typically reduces the detection rate by a factor of $\sim 5$ at an $OFAP$ of $10\%$.

\subsection{Tuning HACR for multiple classes of source}
\label{multtune}
\begin{figure}[!hb]
\begin{center}
\includegraphics[height=5in,angle=0]{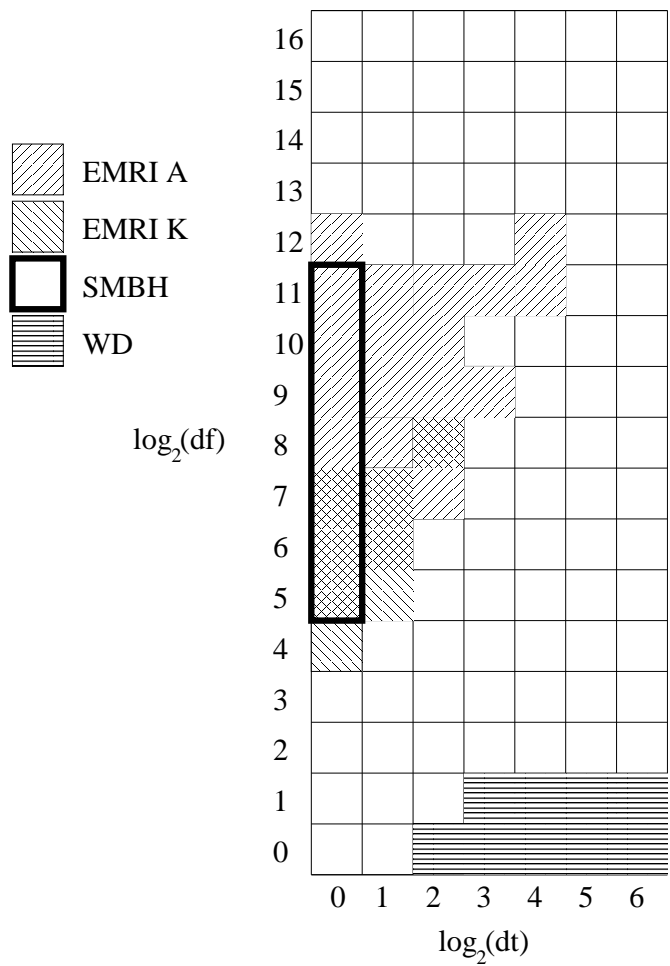}
\end{center}
\caption{Box sizes in which the majority of detections are made for various sources. For each of four different sources --- EMRI ``A'' at 2Gpc, EMRI ``K'' at 2Gpc, the SMBH binary at z=3.125 and the WD binary at 17kpc --- we indicate all box sizes which were responsible for $>1\%$ of the detections of that source in 1000 realisations. The sources are identified by the patterns in the key. Box sizes that were good for several sources are indicated by multiple patterns, e.g. the box with co-ordinates (0,7).}
\label{BestBin}
\end{figure}

One solution to this problem in a LISA search would be to run several independent searches focussed on different source families. However, it is also possible to tune a single set of HACR thresholds to be sensitive to all three types of source simultaneously. This is done in the same way as the source and distance-averaged tuning described in Section~\ref{multEMRItune}, but now we inject not only EMRI signals, but also WD and SMBH signals. 
When the thresholds are tuned using EMRI source ``A'' at 2Gpc, the WD binary at 17kpc and the SMBH binary at $z=3.125$ with equal weighting, 
the detection rate at an $OFAP$ of $10\%$ for the EMRI source ``A''  at 2Gpc is $87.0\%$ as opposed to $89.3\%$ using optimal tuning. 
This difference is of the same order as the expected error in our ROC estimates (see Section~\ref{HACRtune}) and is therefore considered to be negligible. 
For the SMBH binary at $z=3.125$ and the WD binary at 17kpc the change in detection rate when using the thresholds tuned for all 
three sources when compared to th detection rate obtained using the optimal thresholds is also negligible. 
It is clear that when the thresholds are tuned for all three types of source, the performance of HACR is almost as high as the source specific searches, and still exceeds the performance of the Excess Power search. That this is possible follows from the different time-frequency properties of the three types of source. The time-frequency properties of a source determine which box sizes are good for its detection. This is illustrated in Figure~\ref{BestBin}, which shows schematically all box sizes that contribute more than $1\%$ of the detection rate for four different sources: EMRIs ``A'' and ``K'', the WD binary and the SMBH binary inspiral. Physically, we expect WD binary tracks to be virtually monochromatic, and of long duration. Therefore
we might expect to detect such sources in box sizes that are long in time but very narrow in frequency. The SMBH binary inspiral (at that redshift) is fairly short in duration, but sweeps through a reasonable range in frequency and is also quite
loud. Therefore, we might expect to see it in boxes that are narrow in time, and
broader in frequency. EMRIs are similar in structure to SMBH binary inspirals, but last longer in time and evolve more slowly. For a circular EMRI (e.g., source ``K''), one might expect to detect it in boxes that were long in time and quite narrow in frequency, although shorter in time and slightly broader in frequency than the WD binary (since the frequency changes as the source inspirals). However, an eccentric EMRI (e.g., source ``A'') will have multiple frequency harmonics, and one might expect to do better using a slightly broader box in frequency which then includes more of the frequency components. The distribution in Figure~\ref{BestBin} fits precisely with this physical intuition. When tuning for multiple sources, the threshold in a given box size will be determined by the source that the box size is most suited to detecting. The fact that the various types of source favour distinct groups of box sizes means the overall performance is comparable to the source specific performance. The box sizes in which HACR detections are made thus provide an additional way to classify the source type.

\section{Using HACR for parameter estimation}
\label{paramest}
We have emphasised throughout this paper that, although the HACR search does not provide a much greater detection rate than the Excess Power search, the clusters it identifies may be used to characterize the source. Parameters estimated from HACR clusters can then be used as input for other algorithms in subsequent stages of a hierarchical search of the LISA data. An Excess Power detection essentially contains two pieces of information --- the time and frequency at which the detection is made. Since we are using binning as part of the search, there is also some information contained in the box size(s) used to bin the spectrogram(s) in which the detection(s) is (are) made. To gain further information, a detection made by Excess Power must be followed by a track identification stage, and this is currently being investigated \cite{wen06}. A HACR cluster by contrast consists not of one but many pixels. Thus, in addition to the previous properties, the HACR cluster has shape information which is potentially a much more powerful diagnostic. The information that can be extracted ranges from the size of the event in time and frequency to the shape and curvature of the boundary of the cluster. An event that is short in the time direction but broad in frequency might be an instrumental noise burst, whereas events long in time and narrow in frequency are probably inspiral events. The difference in frequency between the latest and earliest pixels in the cluster, divided by the difference in time, provides an estimate of the rate of change of frequency (or chirp rate) of the event. The shape parameters \cite{sahni98} of the cluster also provide diagnostics which might be able to distinguish instrumental bursts from astrophysical bursts from long lived astrophysical events. As mentioned elsewhere, source confusion is a major issue for LISA, with many events likely to be overlapping in time and frequency in the data stream. A detection in the time-frequency plane could therefore either be a single source or several overlapping sources. An analysis of the cluster boundary should be able to distinguish these two cases in certain situations, i.e., distinguish a ``cross'' from a ``line''.

The power profile in the cluster is also a potentially useful diagnostic. One use would be to distinguish crossing tracks from inspirals as above. Additionally, the power profile along an inspiral track would reveal the modulations associated with the motion of the detector and thus provide a diagnostic of sky position. In a more sophisticated analysis, cluster properties would allow different clusters that are generated by the same event to be identified. An EMRI is characterized by several different frequency components and these might well appear as different clusters in a time-frequency analysis (see spectrograms in \cite{wengair05}). However, these tracks remain almost parallel as they evolve, and so the rate of change of frequency provides a way to connect the tracks in a second stage analysis of the HACR clusters. If tracks can be identified like this, the properties, such as the track separation, encode information about the orbital eccentricity etc.

One complication in all of this is that the construction of the binned spectrograms makes use of bins that overlap in time and frequency. This has the effect of smearing out tracks from astrophysical sources and noise events in the data, which complicates cluster characterisation and parameter extraction. In analysing cluster properties, this effect must be accounted for, or methods derived to deconvolve the effect of binning once a source has been identified.

It is clear that HACR cluster properties are a potentially powerful tool both for vetoing, i.e., distinguishing astrophysical events from instrumental artefacts, and for parameter estimation. Work is currently underway to investigate which of these and other cluster properties are most powerful as diagnostics, and how the system's parameters may be estimated from them. However, we leave a fuller discussion of this analysis and the results for a future paper.


\section{Concluding Remarks}
\label{conclusion}
In this paper we have described a time-frequency algorithm, the Hierarchical Algorithm for Clusters and Ridges, that can be used to detect gravitational wave signals in LISA data. The algorithm extends the simple Excess Power search described elsewhere \cite{wengair05,gairwen05} and is similar to the TFClusters algorithm used for LIGO data analysis \cite{jsylvestre}. We have investigated how the thresholds in the HACR algorithm may be tuned for specific sources and examined the performance of the algorithm for detection of three expected LISA sources --- extreme mass ratio inspirals, white dwarf binaries and supermassive black hole mergers. Our results suggest that the algorithm can detect typical EMRI events at distances of up to $\sim 2.6$Gpc, typical WD binary events at several kpc (up to $20$kpc for our favourably oriented example) and typical SMBH merger events at up to redshift $z\sim3.5$. Moreover, we have demonstrated that it is possible to tune HACR to be sensitive to all three of these distinct waveform families simultaneously. This is possible because the time-frequency structure of the sources is quite different. A key ingredient of the search (as in the Excess Power search) is to bin the data using boxes of certain sizes. The time-frequency structure of the waveform family determines which boxes are well-suited to their detection, and these sets of box sizes are largely distinct for the different waveform types. This allows the overall search to be tuned for all three source families.

The HACR search includes the Excess Power search as a special case (i.e., when the pixel threshold is set equal to one). HACR must therefore perform at least as well as the Excess Power search. In fact, we find that for the detection of a single source HACR has a detection rate about $5\%-20\%$ better than the Excess Power search. This may seem like a modest improvement, but it translates to a fairly significant enhancement in event rate. For sources at distances close to the detection limit, the HACR and Excess Power searches have very similar performance. Moreover, HACR represents an improvement over Excess Power since the HACR events are clusters containing several hundred pixels, rather than the single pixel identified in the Excess Power search. Parameter extraction from an Excess Power search requires a follow up stage of track identification \cite{wen06}. The HACR pixel clusters, on the other hand, already directly encode information about the type of source and the source parameters. We have discussed some ways in which information can be extracted from the cluster and mapped into physical parameters, but we reserve a more in-depth discussion of this important part of the search for a future paper.

The current work contains some limitations which will also be explored in the future. The exploration of the waveform parameter space has been far from exhaustive --- we have considered only a few EMRI sources, plus single examples of signals from WD binaries and SMBH binaries. A more thorough examination of the parameter space is required to fully assess the usefulness of HACR in terms of likely LISA event rates. Our model for the LISA data stream is also somewhat simplified --- we have used a low-frequency approximation, rather than the full TDI description of the detector output. While these approximations should not seriously change the conclusions, it will be important to model LISA more accurately in future studies. The most significant limitation in the current work is the fact that we have considered the extraction of a single event from noisy data. In  practice, the LISA data stream will contain many thousands of events overlapping in time and frequency. The performance of HACR and other techniques will be very different under those circumstances than under the ideal conditions considered here. However, the most likely effect of source confusion will just be to limit the distance to which sources can be seen. We should still be able to identify the loudest handful of events using this or more sophisticated techniques, at considerably lower computational cost than many other approaches. The cluster properties will also provide additional information to help disentangle multiple crossing tracks. The performance of HACR in a source-dominated data stream will also be explored in the future. In addition, we plan to assess HACR by analysing the LISA Mock Data Challenge data sets \cite{mldc}.

In conclusion, we hope that HACR, or some similar time-frequency technique, will be able to provide a computationally cheap first stage in a hierarchical search of LISA data. The parameter estimates obtained from the clusters can the be used as input for subsequent follow up with matched filtering or Markov Chain Monte Carlo techniques. This should allow detection of the loudest $\sim 10$s of LISA events for a very low computational overhead.

\acknowledgments
We thank Stanislav Babak, Bangalore Sathyaprakash and R Balasubramanian for useful discussions. This work was supported by St. Catharine's College, Cambridge (JG) 
and by the School of Physics and Astronomy, Cardiff University (GJ).

\end{document}